\documentstyle{aipproc}
\begin{document}
\title{Anharmonic phonon excitations \\ in subbarrier fusion reactions}
\author{K. Hagino$^{*}$, N. Takigawa$^{*}$, and S.  Kuyucak$^{\dagger}$}
\address{$^{*}$Department of Physics,
Tohoku University, Sendai 980--77, Japan
\\
$^{\dagger}$Department of Theoretical Physics, Research School 
of Physical Sciences, \\ Australian National University, 
Canberra, ACT 0200, Australia
}

\maketitle

\begin{abstract}
Recently measured high precision data of fusion 
excitation function 
have enabled a detailed study on the effects of nuclear collective 
excitations on fusion reactions. Using such highly accurate data 
of the $^{16}$O + $^{144,148}$Sm reactions, we discuss the anharmonic 
properties of collective phonon excitations in $^{144,148}$Sm nuclei. 
It is shown that subbarrier fusion reactions are strongly affected 
by the anharmonic effects and thus offer an alternative method to 
extract the static quadrupole moments of phonon 
states in a spherical nucleus. 

\end{abstract}

\section*{INTRODUCTION}

It has been well recognized that cross sections of heavy-ion 
fusion reactions at energies near and below the Coulomb barrier 
are strongly influenced by couplings of the relative motion of 
the colliding nuclei to several nuclear intrinsic motions \cite{BT97}. 
In the eigen-channel approach, such couplings give rise to a distribution 
of potential barriers \cite{HTB97,DLW83}. 
Based on this idea, a method was proposed to extract barrier 
distributions directly from fusion excitation functions using 
the second derivative of the product of the fusion cross section 
and the center of mass energy $E \sigma$ as a function of 
energy $E$ \cite{RSS91}. 
Based on coupled-channels calculations, it was shown that the fusion 
barrier distribution, i.e. $d^2(E\sigma)/dE^2$, is very sensitive to 
the details of the couplings. 
In order to deduce meaningful barrier distributions, 
excitation functions of fusion cross sections have to be 
measured with high precision at small energy intervals. 
Thanks to the recent 
developments in experimental techniques \cite{WLH91}, such data are now 
available for several systems, and they have clearly demonstrated 
that the barrier distribution is indeed a sensitive quantity 
to channels couplings\cite{LDH95}.

In this contribution, we analyse the recently measured accurate 
data on the $^{16}$O + $^{144}$Sm fusion reaction to discuss effects of 
nuclear surface vibrations on heavy-ion fusion reactions \cite{HTK97}. 
The barrier distribution analysis of the recent high precision 
data on the $^{58}$Ni + $^{60}$Ni fusion reaction has shown clear 
evidence for coupling of multi-phonon states in 
$^{58}$Ni and $^{60}$Ni \cite{SACN95}, while no evidence for 
double phonon couplings is seen in the $^{16}$O + $^{144}$Sm 
reaction \cite{LDH95,MDH94}. 
We show that anharmonicities in nuclear vibrations 
play an important role in the latter reaction. 
We estimate the magnitude as well as the sign of the quadrupole 
moments of the quadrupole and octupole 
single-phonon states of $^{144}$Sm from the experimental fusion barrier 
distribution. A similar analysis is performed also for the
$^{16}$O + $^{148}$Sm reaction. 

\section*{ANHARMONICITIES IN NUCLEAR VIBRATIONS}

Collective phonon excitations are common phenomena in fermionic many-body
systems.  In nuclei, low-lying surface oscillations with various
multipolarities are typical examples.  The harmonic vibrator provides a 
zeroth order description for these surface oscillations, dictating simple 
relations among the level energies and the electromagnetic transitions 
between them.  For example, all the levels in a phonon multiplet are 
degenerate and the energy spacing between neighboring multiplets is a 
constant.  In realistic nuclei, however, there are residual interactions 
which cause deviations from the harmonic limit, e.g., they split levels 
within a multiplet, change the energy spacings, and also modify the ratios 
between various electromagnetic transition strengths.  There are many 
examples of two-phonon triplets ($0^+,2^+,4^+$) of quadrupole surface 
vibrations in even-even nuclei near closed shells.  Though the center of 
mass of their excitation energies are approximately twice the energy of the 
first $2^+$ state, they usually exhibit appreciable splitting within the 
multiplet.  A theoretical analysis of the anharmonicities for the 
quadrupole vibrations was first performed by Brink {\it et al.} 
\cite{BTK65}, where they related the excitation energies of three-phonon 
states to those of double-phonon triplets.  For a long time, however, the 
sparse experimental data on three-phonon states had caused debates on the 
existence of multi-phonon states.  The experimental situation has improved 
rapidly in recent years, and data on multi-phonon states are now available 
for several nuclei. 
As a consequence, study of multi-phonon states, and especially their 
anharmonic properties, is attracting much interest \cite{CZ96}. 
It is worthwhile to mention that anharmonic effects are not restricted to 
low-lying vibrations but have also been observed in multi-phonon 
excitations of giant resonances in heavy-ion collisions \cite{LAC97}.  

In many even-even nuclei near closed shells, a low-lying 3$^{-}$ excitation 
is observed at a relatively low excitation energy, which competes with the 
quadrupole mode of excitation \cite{BM75}.  These excitations have been 
frequently interpreted as collective octupole vibrations arising from a 
coherent sum of one-particle one-hole excitations between single particle 
orbitals differing by three units of orbital angular momentum.  This 
picture is supported by large E3 transition probabilities from the first 
3$^{-}$ state to the ground state, and suggests the possibility of 
multi-octupole-phonon excitations.  In contrast to the quadrupole 
vibrations, however, so far there is little experimental evidence for 
double-octupole-phonon states.  One reason for this is that E3 transitions 
from two-phonon states to a single-phonon state compete against E1 
transitions.  This makes it difficult to unambiguously identify the 
two-phonon quartet states ($0^+,2^+,4^+,6^+$).  Only in recent years, 
convincing evidences have been reported for double-octupole-phonon states in 
some nuclei, including $^{208}$Pb \cite{YGM96} and $^{144}$Sm \cite{GVB90}. 

\section*{EFFECTS OF PHONON EXCITATIONS ON FUSION}

Let us now discuss the effects of nuclear surface vibrations on 
heavy-ion fusion reactions. In this section we use the linear 
coupling approximation to describe the coupling between the 
relative motion of the colliding nuclei and the surface 
vibrations. This simple model enables us to understand easily 
the effects of anharmonicity. Extension of the model so as 
to include the couplings to all orders and comparisons with the 
experimental data is given in the next section. 

\subsection*{Harmonic limit}

The effects of nuclear surface vibrations on 
heavy-ion fusion reactions at energies below and near the 
Coulomb barrier has been investigated by many groups 
(see Ref.~\cite{BT97} for a recent review). 
These studies were later extended to include the effect of
multi-phonon states within the harmonic oscillator 
approximation \cite{TI86,KRNR93}. 
Using the no-Coriolis approximation \cite{TI86} and the 
linear coupling approximation, the coupling Hamiltonian, which 
describes the coupling between the relative motion and the quadrupole 
surface oscillations, is assumed to be 
\begin{equation}
V_{coup}(r,\xi)=\frac{\beta}{\sqrt{4\pi}}f(r) (a_{20}^{\dagger} 
+ a_{20}), 
\end{equation}
where $a_{20}^{\dagger}$ and $a_{20}$ are the creation and the 
annhilation operators for the quadrupole phonon, respectively, and 
$\beta$ is the quadrupole deformation parameter. 
The coupling form factor $f(r)$ consists of the nuclear and 
Coulomb parts and reads
\begin{equation}
f(r)=-R_T\frac{dV_N}{dr}+\frac{3}{5}Z_PZ_Te^2\frac{R_T^2}{r^3}. 
\end{equation}
Here $R_T$ is the radius of the target nucleus and $V_N$ is the nuclear 
potential. 

For the quadrupole surface vibrations, the two phonon state has three 
levels ($0^+,2^+,4^+$). 
In the harmonic limit, this two-phonon triplet 
is degenerate in the excitation energy. 
One can then introduce the 
two-phonon channel by taking particular linear combinations of the wave 
functions of the two-phonon triplet \cite{TI86,KRNR93}. 
The wave function of the two-phonon channel then reads
\begin{equation}
|2> = \sum_{I=0,2,4} <2020|I0> |I0> = 
\frac{1}{\sqrt{2!}}(a^{\dagger}_{20})^2 |0>. 
\end{equation}
In the same way, one can introduce the $n$-phonon channel as 
\begin{equation}
|n> = \frac{1}{\sqrt{n!}}(a^{\dagger}_{20})^n |0>. 
\end{equation}
The dimension of the coupled-channels equations is reduced 
drastically with the introduction of the $n$-phonon channels. 
If we truncate to the two phonon states, the corresponding coupling 
matrix is given by 
\begin{equation}
V_{coup}=
\left(\begin{array}{ccc}
0&F(r)&0\\
F(r)&\hbar\omega
&\sqrt{2}F(r)\\
0&\sqrt{2}F(r)&2\hbar\omega
\end{array}\right).
\end{equation}
Here, $F(r)$ is defined as $\frac{\beta}{\sqrt{4\pi}}f(r)$. 

\subsection*{Anharmonic vibrator}

The $sd$-interacting boson model (IBM) in the vibrational limit provides a
convenient calculational framework to discuss the effects of anharmonicity 
in the surface vibrations \cite{IA87}.
The vibrational limit of the IBM and the anharmonic vibrator (AHV) in the
geometrical model are very similar, the only difference coming from
the finite number of bosons in the former \cite{CW88}. 
A model for subbarrier fusion
reactions, which uses the IBM to describe effects of channel couplings, has
been developed in Ref.~\cite{BBK94}.  Following Ref.~\cite{BBK94}, we assume
that the coupling Hamiltonian is given as 
\begin{equation}
V_{coup}(r,\xi)=\frac{\beta}{\sqrt{4\pi N}}f(r) Q_{20}. 
\end{equation}
Here, $N$ is the boson number and we have introduced the 
scaling of the coupling strength with $\sqrt{N}$ to ensure
the equivalence of the IBM and the geometric model results in the large $N$
limit \cite{BBK94}. $Q_{20}$ is the quadrupole operator in the IBM, 
which we take as 
\begin{equation}
Q_{20}=s^{\dagger}d_0 + sd_0^{\dagger} +
\chi_2(d^{\dagger}\tilde{d})^{(2)}_0,
\end{equation}
where tilde is defined as $\tilde{b}_{l\mu}=(-)^{l+\mu}b_{l-\mu}$.  

As in the harmonic limit, one can introduce the multi-phonon channel 
if one assumes that the multi-phonon multiplets are degenerate in the 
excitation energy. The wave function of the $n$-phonon channel in 
the framework of the IBM then reads 
\begin{equation}
|n>=\frac{1}{\sqrt{n!(N-n)!}}(s^{\dagger})^{N-n}(d_0^{\dagger})^n
|0>.
\end{equation}
The corresponding coupling matrix, truncated to the two-phonon states, 
is given by 
\begin{equation}
V_{coup}=
\left(\begin{array}{ccc}
0&F(r)&0\\
F(r)&\hbar\omega-\frac{2}{\sqrt{14N}}\chi F(r)
&\sqrt{2(1-1/N)}F(r)\\
0&\sqrt{2(1-1/N)}F(r)&2\hbar\omega+\delta
-\frac{4}{\sqrt{14N}}\chi F(r)
\end{array}\right).
\end{equation}
The parameter $\delta $ is introduced to represent deviation of 
the energy spectrum from the harmonic limit. 
When the $\chi$ parameter in the quadrupole operator is zero, 
quadrupole moments of all states vanish, and one obtains the harmonic 
limit in the large $N$ limit. Non-zero values of $\chi$ generate 
quadrupole moments and, together 
with finite boson number, they are responsible for the anharmonicities in 
electric transitions.

It has been shown that anharmonicities in level energies have
only a marginal effect on the fusion excitation function and the barrier
distribution \cite{HTK97}. In fact, our studies show that the fusion barrier
distribution does not depend so much on the excitation
energies of the multi-phonon states once the energy of the single-phonon 
state is fixed.  
We therefore set $\delta $ to be zero in the following discussion. 
As we will see later, the main
effects of anharmonicity on fusion barrier distributions come from the
deviation of the transition probabilities from the harmonic limit 
as well as the reorientation effects.

\section*{COMPARISON WITH EXPERIMENTAL DATA}

\subsection*{The $^{16}$O + $^{144}$Sm reaction}

We now discuss the effects of anharmonicities on 
the $^{16}$O + $^{144}$Sm fusion reaction, whose excitation function has
recently been measured with high accuracy \cite{LDH95}. 
It has been reported that inclusion of the 
double-phonon excitations of $^{144}$Sm in coupled-channels calculations in 
the harmonic limit destroys the good agreement between the experimental 
fusion barrier distribution and the theoretical predictions obtained when 
only the single-phonon excitations are taken into account \cite{M95}. On 
the other hand, there are experimental \cite{GVB90,WRZB96} as well as 
theoretical \cite{GS94} support for the existence of the 
double-octupole-phonon states in $^{144}$Sm.  Reconciliation of these 
apparently contradictory facts may 
be possible if one includes the 
anharmonic effects, which are inherent in most multi-phonon spectra.

In order to address these questions, it is necessary to 
extend the models which were discussed in the previous section so that they 
include the octupole mode as well as the couplings to all orders. 
The full order treatment is crucial in order to quantitatively, 
as well as qualitatively, 
describe heavy-ion subbarrier fusion reactions \cite{BBK94,HTDHL97a}. 
We therefore assume the following coupling Hamiltonian based on the 
$sdf$-IBM. 
\begin{eqnarray}
&& V_{coup}(r,\xi)=V_C(r,\xi) + V_N(r,\xi), \\
&&V_C(r,\xi)=\frac{Z_PZ_Te^2}{r}\left(1
+ \frac{3}{5}\frac{R_T^2}{r^2} \frac{\beta_2 \hat{Q}_{20}}{\sqrt{4\pi N}}
+ \frac{3}{7}\frac{R_T^3}{r^3} \frac{\beta_3 \hat{Q}_{30}}{\sqrt{4\pi N}}
\right) , \nonumber \\
&&V_N(r,\xi)=-V_0 \left[ 1+\exp \left(\frac{1}{a} \left( r-R_0 - R_T (
\beta_2 \hat{Q}_{20} + \beta_3 \hat{Q}_{30})/ \sqrt{4\pi N} \right)
\right)\right]^{-1}.
\label{v}
\end{eqnarray}
The quadrupole and the octupole operators are defined as 
\begin{eqnarray}
&&\hat{Q}_2=s^{\dagger}\tilde{d} + sd^{\dagger} +
\chi_2(d^{\dagger}\tilde{d})^{(2)}
+ \chi_{2f}(f^{\dagger}\tilde{f})^{(2)}, \nonumber \\
&&\hat{Q}_3=sf^{\dagger} +
\chi_3(\tilde{d}f^{\dagger})^{(3)} + h.c.,
\label{op}
\end{eqnarray}
respectively. 

The results of the coupled-channels calculations are compared with the 
experimental data in Fig.~1.  The upper and the lower panels in Fig.~1 show 
the excitation function of the fusion cross section and the fusion barrier 
distributions, respectively. The experimental data are taken from 
Ref.~\cite{LDH95}. 
The dotted line is the result in the harmonic limit, 
where couplings to the quadrupole and octupole vibrations in $^{144}$Sm are 
truncated at the single-phonon levels and all the $\chi$ parameters in 
Eq.~(\ref{op}) are set to zero.  The deformation parameters in 
Eq.~(\ref{v}) are estimated to be $\beta_2$=0.11 and $\beta_3$=0.21 from 
the electric transition probabilities.
The dotted line reproduces the experimental data of both the fusion cross 
section and the fusion barrier distribution reasonably well, though the 
peak position of the fusion barrier distribution around $E_{cm}=65$ MeV is 
slightly shifted.  As was shown in Ref.~\cite{M95}, the shape of the fusion 
barrier distribution becomes inconsistent with the experimental data when 
the double-phonon channels are included in the harmonic limit (the dashed 
line).  
To see whether this discrepancy is due to neglecting of anharmonic 
effects, we have repeated the calculations including the $\chi$ parameters 
in the fits and using $N=2$ in the IBM. 
The $\chi^2$ fit to the fusion cross sections resulted in the set of 
parameters, $\chi_2=-3.30\pm 2.30$, $\chi_{2f}=-2.48\pm 0.07$, and 
$\chi_3=2.87\pm 0.16$, regardless of the starting values. 
The resulting fusion cross sections and barrier distributions are shown in 
Fig.~1 by the solid line.  They agree with the experimental data much 
better than those obtained in the harmonic limit.  Thus, inclusion of the 
anharmonic effects in vibrational motion appear to be essential for a 
proper description of barrier distributions in the reaction $^{16}$O + 
$^{144}$Sm.

One of the pronounced features of an anharmonic vibrator is that the 
excited states have non-zero quadrupole moments \cite{BM75}.  Using the 
$\chi$ parameters extracted from the analysis of fusion data in the E2 
operator, $T$(E2$)=e_B \hat Q_2$, we can estimate the static quadrupole 
moments of various states in $^{144}$Sm.  Here, $e_B$ is the effective 
charge, which is determined from the experimental $B($E2$;0\to 2^+_1)$ 
value as $e_B = 0.16~eb$.  For the quadrupole moment of the first 2$^+$ and 
3$^-$ states, we obtain $-$0.89 $\pm$ 0.63 b and $-$0.70 $\pm$ 0.02 b, 
respectively.  
Fig.~2 shows the
influence of the sign of the quadrupole moment of the excited states on the 
fusion cross section and the fusion barrier distribution.  The solid line 
is the same as in Fig.~1 and corresponds to the optimal choice for the 
signs of the quadrupole moments of the first 2$^+$ and 3$^-$ states.  The 
dotted and dashed lines are obtained by changing the sign of the $\chi_2$ 
and $\chi_{2f}$ parameters in Eq.~(\ref{op}), respectively, while the 
dot-dashed line is the result where the sign of both $\chi_2$ and 
$\chi_{2f}$ parameters are inverted.  The change of sign of $\chi_2$ and 
$\chi_{2f}$ is equivalent to taking the opposite sign for the quadrupole 
moment of the excited states.  
Fig.~2 demonstrates that subbarrier fusion 
reactions are sensitive to the sign of the quadrupole moment of 
excited states.  
The experimental data are reproduced only when the correct 
sign of the quadrupole moment are used in the coupled-channels 
calculations.  

\subsection*{The $^{16}$O + $^{148}$Sm reaction}

A similar analysis was performed also for the $^{16}$O + $^{148}$Sm 
reaction \cite{HTK97b}. 
The best fit to the experimental fusion cross section \cite{LDH95} 
was obtained with the quadrupole moments of 
$-1.00 \pm 0.25 b$ and $+1.52 \pm 0.14 b$ for the first 2$^+$ and 
3$^-$ states, respectively. The total boson number $N$ was assumed 
to be 4 and the deformation parameters were estimated from 
the electric transition probabilities.
Note that the value of the quadrupole moment of the first 2$^+$ state 
which we obtained from the fusion analysis is very close to that 
measured from the Coulomb excitation technique, i.e. 
$-0.97 \pm 0.27 b$ \cite{P90}. 
Fig.~3 shows the 
sensitivity of the fusion cross section and the fusion barrier 
distribution to the sign of the quadrupole moment of the first 
3$^-$ state. The experimental data are taken from Ref.~\cite{LDH95}. 
The solid line corresponds to the optimal choice for the 
sign of the first 3$^-$ state, while the dotted line was 
obtained by inverting it. 
We again observe that the use of the incorrect sign of 
the quadrupole moment destroys the good fit to the experimental 
data. 
This strongly suggests that subbarrier fusion can provide an alternative 
method to determine the sign as well as the magnitude of the 
quadrupole moments in spherical nuclei. 

\section*{SUMMARY}

We discussed the effects of multi-phonon excitations on 
subbarrier fusion reactions. We especially focused on the 
anharmonic properties of the phonon excitations. 
The experimental fusion excitation functions for the 
$^{16}$O + $^{144,148}$Sm reactions were analyzed with a model which 
explicitly takes into account the effects of anharmonicity of 
the vibrational modes of excitation in $^{144,148}$Sm.  
We found that the best fit to the 
experimental data requires negative quadrupole moments for the first 2$^+$
and the first 3$^-$ states of $^{144}$Sm. 
For the $^{148}$Sm nucleus, we obtained a negative quadrupole moment 
for the first 2$^{+}$ state and a positive one 
for the first 3$^{-}$ state. 
As a general conclusion, we find that 
heavy-ion subbarrier fusion reactions, and in particular, barrier 
distributions extracted from the fusion data, are very sensitive to the 
sign of the quadrupole moments of phonon states in the target nucleus 
and thus offer an alternative method to determine them. 

\section*{ACKNOWLEDGMENTS}

The authors thank J.R. Leigh, M. Dasgupta, D.J. Hinde, and J.R. Bennett for
useful discussions.  
The work of K.H. was supported by the Japan Society for the Promotion of
Science for Young Scientists.  
This work was also supported by the Grant-in-Aid for General                
Scientific Research,                                                        
Contract No.08640380, Monbusho International Scientific Research Program:   
Joint Research, Contract No. 09044051,                                      
from the Japanese Ministry of Education, Science and Culture,               
and a bilateral program of JSPS between Japan and Australia.

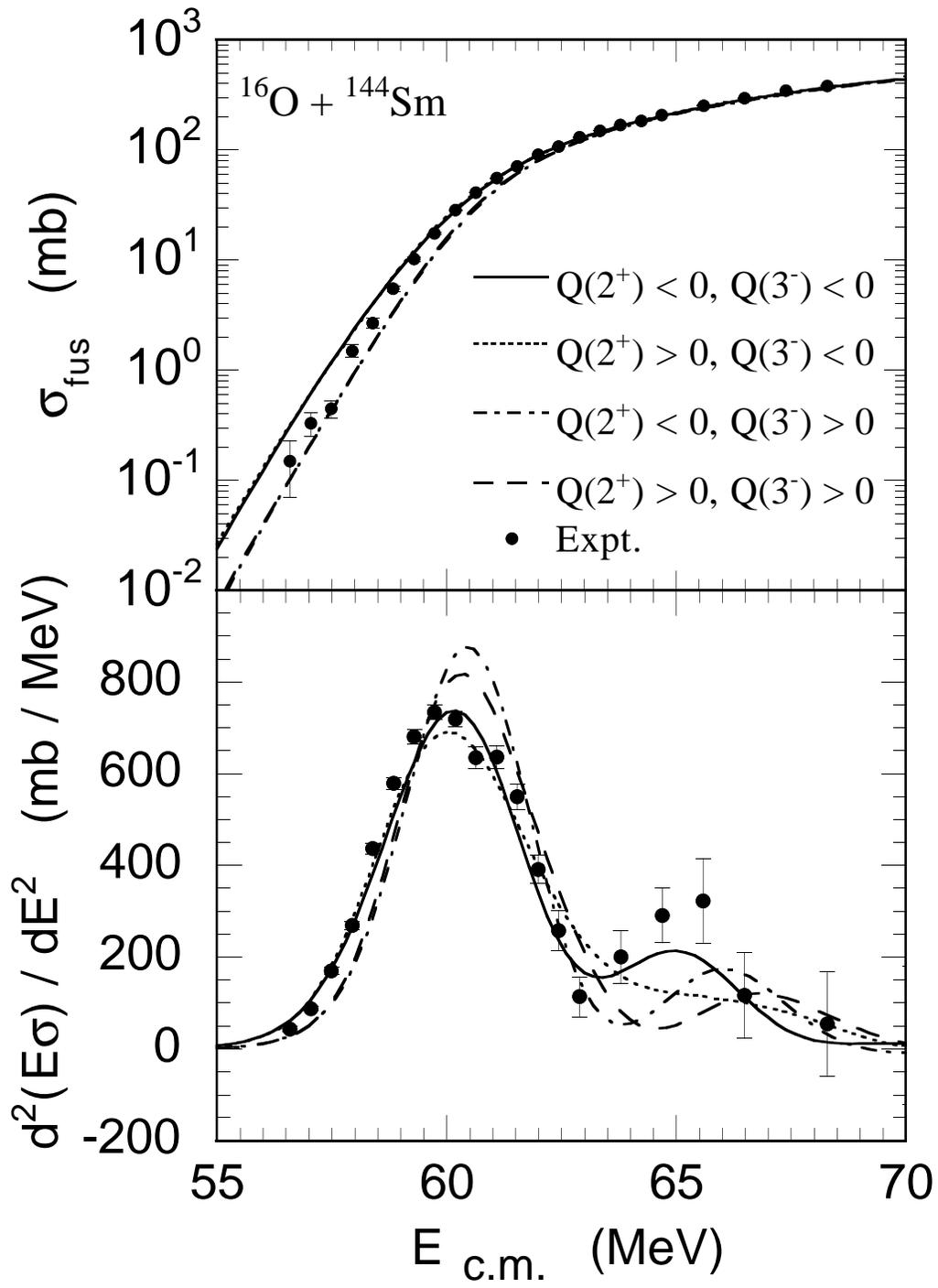
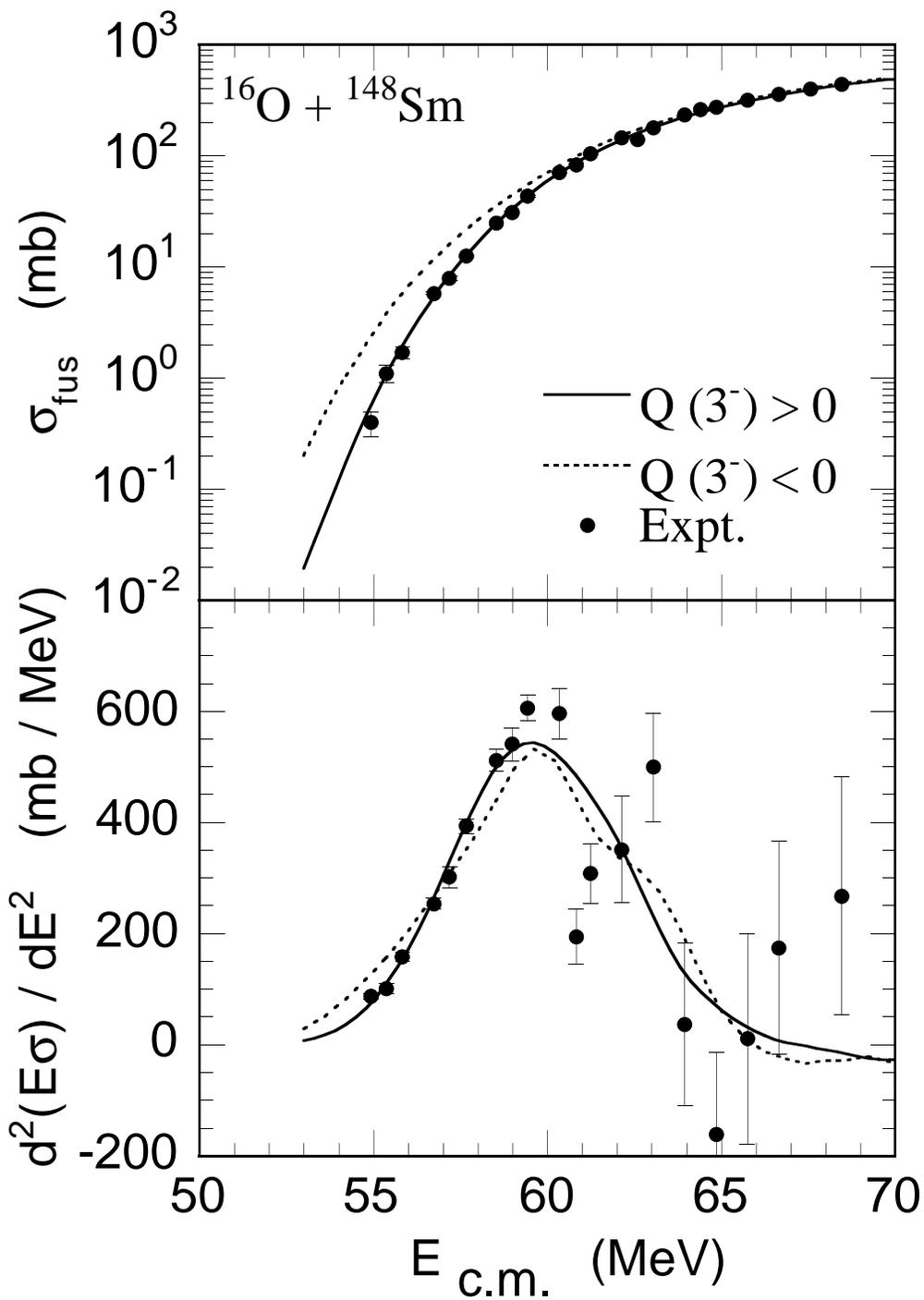
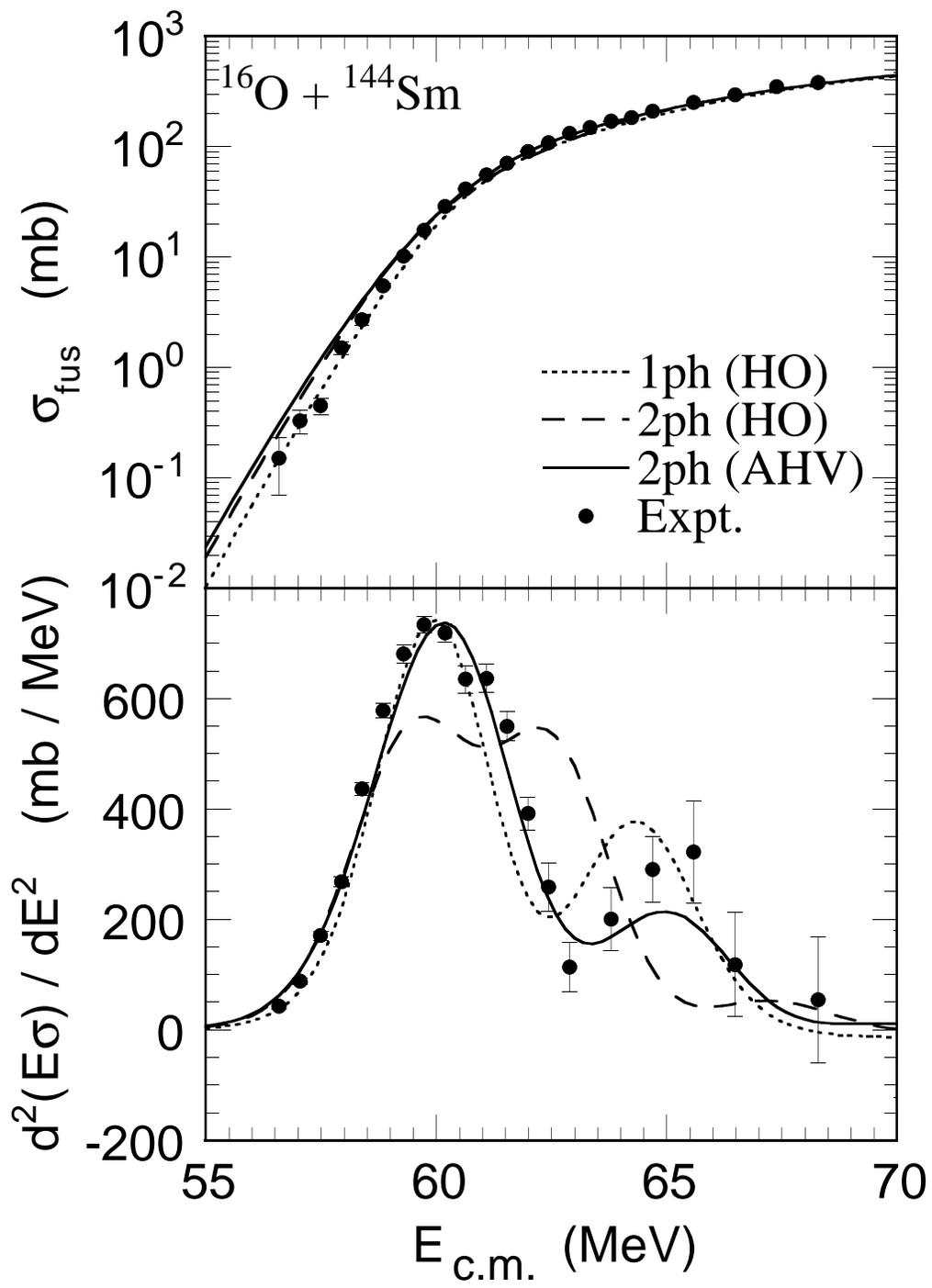
\end{document}